\documentclass[article, aps, prb, twocolumn, superscriptaddress, floatfix, showpacs]{revtex4-1}

\usepackage[utf8]{inputenc}
\usepackage{graphicx}
\usepackage{mathtools}
\usepackage{braket}
\usepackage{hyperref}
\usepackage{amssymb}
\graphicspath{{./pictures/}}

\usepackage[T2A]{fontenc}
\ifpdf\usepackage{epstopdf}\fi

\graphicspath{{pictures/}}

\begin{document}

\author{A. I. Pakhomchik}
\affiliation{Moscow Institute of Physics and Technology, Institutskii Per. 9, Dolgoprudny, Moscow Region 141700, Russia}
\author{I. Feshchenko}
\affiliation{Skolkovo Institute of Science and Technology,
5 Nobelya str., Mozhajskij region, Moscow, 121205, Russia}
\affiliation{Moscow Institute of Physics and Technology, Institutskii Per. 9, Dolgoprudny, Moscow Region 141700, Russia}
\author{A. Glatz}
\affiliation{Materials Science Division, Argonne National Laboratory, 9700  S. Cass  Avenue,  Argonne,  IL  60439,  USA}
\author{V. M.  Vinokur}
\affiliation{Materials Science Division, Argonne National Laboratory, 9700  S. Cass  Avenue,  Argonne,  IL  60439,  USA}
\author{A. V. Lebedev}
\affiliation{Moscow Institute of Physics and Technology, Institutskii Per. 9, Dolgoprudny, Moscow Region 141700, Russia}
\author{K. V. Kuzhamuratova}
\affiliation{Moscow Institute of Physics and Technology, Institutskii Per. 9, Dolgoprudny, Moscow Region 141700, Russia}
\author{S. N. Filippov}
\affiliation{Moscow Institute of Physics and Technology, Institutskii Per. 9, Dolgoprudny, Moscow Region 141700, Russia}
\affiliation{Steklov Mathematical Institute of Russian Academy of Sciences, Gubkina St. 8, Moscow 119991, Russia}
\affiliation{Valiev Institute of Physics and Technology of Russian Academy of Sciences, Nakhimovskii Pr. 34, Moscow 117218, Russia}
\author{G. B. Lesovik}
\affiliation{Moscow Institute of Physics and Technology, Institutskii Per. 9, Dolgoprudny, Moscow Region 141700, Russia}

\title{
    Realization of the Werner-Holevo and Landau-Streater quantum channels for qutrits on quantum computers
}

\begin{abstract}
    We realize Landau-Streater (LS) and Werner-Holevo (WH) quantum channels for qutrits on the IBM quantum computers. These channels correspond to interaction between the qutrit and its environment that result in the globally unitarily covariant qutrit transformation violating multiplicativity of the maximal $p$-norm. Our realization of LS and WH channels is based on embedding qutrit states into states of two qubits and using single-qubit and two-qubit CNOT gates to implement the specific interaction. We employ the standard quantum gates hence the developed algorithm suits any quantum computer. We run our algorithm on a 5-qubit and a 20-qubit computer as well as on a simulator. We quantify the quality of the implemented channels comparing their action on different input states with theoretical predictions. The overall efficiency is quantified by fidelity between the theoretical and experimental Choi states implemented on the 20-qubit computer.
\end{abstract}

\maketitle

\section{Introduction}

A quantum channel (QC) is a completely positive trace-preserving
(CPTP) map, $\Phi$, between operators $\mathcal{B}(\mathcal{H}_d)$
defined on the Hilbert space $\mathcal{H}_d$. Quantum channels
conveniently describe the transformation of a density matrix,
$\rho$, interacting with an
environment\cite{Nielsen_Chuang,breuer-petruccione-2002,Holevo,filippov2019}.
Unitary evolution of the closed system $\rho \rightarrow U \rho
U^{\dagger}$ is an example of the QC that preserves the maximally
mixed state $({1}/{d}) \mathrm{I} $, i.e., $\mathrm{\Phi}
(\frac{1}{d} \mathrm{I}) = \frac{1}{d} \mathrm{I}$. It is
shown\cite{LS} that any open quantum dynamics of a qubit that
preserves the maximally mixed state is an essentially random
unitary one, $\rho \rightarrow \sum_i p_i U_i \rho U_i^{\dagger}$,
however, in larger dimensions this is not the case. Namely, for a
qutrit (d=3) there exists a quantum channel $\mathrm{\Phi}$ such
that $\mathrm{\Phi} (\frac{1}{d} \mathrm{I}) = \frac{1}{d}
\mathrm{I}$ but $\mathrm{\Phi}$ is not random unitary
\cite{LS,Audenaert,Mendl}, and those are just the LS\,\cite{LS}
and WH\,\cite{WH} channels
 that satisfy this condition.
Furthermore, these qutrit channels are extremal, exhibit the
global unitary covariance, and violate the multiplicativity of the
maximal $p$-norm (see\,\cite{WH} and\,\cite{Kuzhamuratova} for a
review). Landau-Streater and Werner-Holevo channels hold high
potential for applications in quantum information sciences. In
this paper, we follow the experimental study of QC on quantum
computers
\cite{Santos2016,AlvarezRodriguez2018,Zhukov2018,Roffe2018,Geller2018,bogdanov2018,Morris}
and implement the Landau-Streater and Werner-Holevo quantum
channels for qutrits on IBM quantum computers with 5 and 20
qubits.

We realize the LS and WH channels using only one-qubit and CNOT
gates. As shown in\,\cite{Nielsen_Chuang}, any unitary operation
can be approximated using these gates up to arbitrary accuracy. We
represent a qutrit, a quantum state being equivalent to a particle
with spin $s=1$, as two qubits ignoring the highest energy level $
\ket{0}_3 \rightarrow \ket{00}, \ket{1}_3 \rightarrow \ket{01},
\ket{2}_3 \rightarrow \ket{10}$. Since the used quantum computers,
\href{https://quantumexperience.ng.bluemix.net}{IBM QUANTUM
EXPERIENCE}\cite{ibm}, are not perfect and are subject to a noise,
the final density matrix may have a non-zero probability of being
in the ignored state. For example, a density matrix can transform
into the following one: $ \rho = (1-\epsilon)
\rho_{\mathrm{qutrit}} + \epsilon (\ket{11} \bra{11} + \ket{11}
\bra{10} + h.c.)$. We discard the last term and normalize $\rho$
with respect to its trace , i.e., we consider only a qutrit part
of the density matrix ($ \rho \rightarrow
\rho_{\mathrm{qutrit}}$). We construct the Choi matrix
corresponding to the QCs from the output density matrices using
different input states. According to the Choi-Jamio{\l}kowski
isomorphism~\cite{choi-1975,jamiolkowski-1972}, the Choi matrix
contains all information about the considered channel. We compare
the obtained matrices with the theoretical expectations obtained
by using a simulator of an ideal quantum computer, which is free
of measurement errors and other types of errors related to the
coupling to the environment. Finally, as a main result, we find
that all the tomography experiments for the Werner-Holevo
$\Phi_{WH}[\rho]$ and Landau-Streater $\Phi_{LS}[\rho]$ channels
agree with the theoretical expectations.

\section{Brief description of the Landau-Streater channel}\label{Steinspring_representation}
The Landau-Streater channel $\Phi$ acts on on the state $\rho$ as follows
\begin{equation}
    \Phi[\rho] = \frac{1}{s(s+1)}(J_x{\rho}J_x+J_y{\rho}J_y+J_z{\rho}J_z),
    \label{LS}
\end{equation}
and is defined through the $SU(2)$ generators $J_x, J_y, J_z$
acting on a $(2s+ 1)$-dimensional Hilbert space
$\mathcal{H}_{2s+1}$ for a spin-$s$ particle. In the case of a
qutrit, $s = 1$, we define the generators
as~\cite{filippovmagadov2018}:
\begin{equation}
\begin{gathered}
    J_x =
    \begin{pmatrix}
        0 & \frac{1}{\sqrt{2}} & 0 \\
        \frac{1}{\sqrt{2}} & 0 & \frac{1}{\sqrt{2}}\\
        0 & \frac{1}{\sqrt{2}} & 0
    \end{pmatrix};
    J_y =
    \begin{pmatrix}
        0 & -\frac{i}{\sqrt{2}} & 0 \\
        \frac{i}{\sqrt{2}} & 0 & -\frac{i}{\sqrt{2}}\\
        0 & \frac{i}{\sqrt{2}} & 0
    \end{pmatrix}; \\
    J_z =
    \begin{pmatrix}
        1 & 0 & 0 \\
        0 & 0 & 0\\
        0 & 0 & -1
    \end{pmatrix}.
\end{gathered}
\end{equation}
In\,\cite{Kuzhamuratova}, the Stinespring
representation for the LS channel was derived and
 relevant matrices $\rho_{\mathrm{env}} \in \mathcal{B}(\mathcal{H}_{\mathrm{env}})$ and
$U, \mathcal{H} \otimes \mathcal{H}_{\mathrm{env}} \rightarrow \mathcal{H} \otimes \mathcal{H}_{\mathrm{env}}$, were calculated and shown to
satisfy the following equation:
\begin{equation}
    \Phi[\rho] = \operatorname{Tr}_{\mathrm{env}}(U(\rho\otimes\rho_{\mathrm{env}})U^{\dagger})\,,
\end{equation}
yielding
\begin{equation}
\rho_{\mathrm{env}} = \ket{0}\bra{0}
\end{equation}\label{Ksenia_env}
and
\begin{equation}
    U =
    \begin{pmatrix}
    \begin{tabular}{c|c|c}
    \begin{tabular}{ccc}
    0 & 0 & 0\\
    0 & 0 & $\frac{i}{\sqrt{2}}$\\
    $\frac{1}{\sqrt{2}}$ & 0 & 0
    \end{tabular} &
    \begin{tabular}{ccc}
    $\frac{1}{2}$ & 0 & 0\\
    $-\frac{i}{2}$ & 0 & 0\\
    0 & $\frac{i}{2}-\frac{1}{2\sqrt{2}}$ & $-\frac{i}{2\sqrt{2}}$
    \end{tabular} &
    \begin{tabular}{ccc}
    0 & $\frac{i}{\sqrt{2}}$ & $-\frac{i}{2}$\\
    0 & 0 & $\frac{1}{2}$\\
    0 & 0 & 0
    \end{tabular}\\ \hline
    \begin{tabular}{ccc}
    $\frac{1}{2}$ & 0 & 0\\
    $\frac{i}{2}$ & 0 & 0\\
    0 & 1 & 0
    \end{tabular} &
    \begin{tabular}{ccc}
    0 & $\frac{1}{2}-\frac{i}{2\sqrt{2}}$ & $\frac{1}{2\sqrt{2}}$\\
    0 & $\frac{1}{2\sqrt{2}}$ & $-\frac{1}{2}-\frac{i}{2\sqrt{2}}$\\
    0 & 0 & 0
    \end{tabular} &
    \begin{tabular}{ccc}
    $\frac{1}{2}$ & 0 & 0\\
    $-\frac{i}{2}$ & 0 & 0\\
    0 & 0 & 0
    \end{tabular}\\ \hline
    \begin{tabular}{ccc}
    0 & 0 & $\frac{1}{\sqrt{2}}$\\
    0 & 0 & 0\\
    0 & 0 & 0
    \end{tabular} &
    \begin{tabular}{ccc}
    $\frac{1}{2}$ & 0 & 0\\
    $\frac{i}{2}$ & 0 & 0\\
    0 & $\frac{1}{2\sqrt{2}}$& $\frac{1}{2}-\frac{i}{2\sqrt{2}}$
    \end{tabular} &
    \begin{tabular}{ccc}
    0 & 0 & $\frac{i}{2}$\\
    0 & $\frac{1}{\sqrt{2}}$ & $\frac{1}{2}$\\
    $-\frac{1}{\sqrt{2}}$ & 0 & 0
    \end{tabular}
    \end{tabular}
    \end{pmatrix}\,.
    \label{Ksenia_matrix}
\end{equation}

\section{Emulation of qutrits by qubits}\label{qutrit_emulation}
The Kraus rank of the LS channel is 3, so one needs at least the three-dimensional environment.
The Hilbert space of two qutrits $\mathcal{H} \otimes \mathcal{H}_{env} = \mathcal{H}_{2s+1} \otimes \mathcal{H}_{2s+1}$ (where $s=1$)
has $3^2$ basis states and the Hilbert space of $N$ qubits has $2^N$ basis states.
Therefore, one needs $N=4$ qubits for the 9 states of two qutrits, which are sufficient to encode a system qutrit that is transformed by the channel and a separate environment qutrit.
In other words, one qutrit uses one pair of qubits, and the other qutrit uses another pair.

For encoding three logical states $m_s=-1$, $m_s=0$, and $m_s=1$
of a qutrit we use the states $\ket{00}$, $\ket{01}$, and
$\ket{10}$, respectively. This choice is related to the fact that
the quantum computer is not perfect, meaning that excited states
may over time evolve back into the ground states due to the
amplitude-damping noise. We did not use physical two-qubit state
$\ket{11}$, because it has the highest energy and, therefore, the
shortest relaxation time into other states. It can decay into the
two states, $\ket{01}$ and $\ket{10}$, which in their turn can
only return to ground state $\ket{00}$. In summary, the algorithm
that transforms $\{\ket{00}, \ket{01},
\ket{10}\}\otimes\{\ket{00}, \ket{01}, \ket{10}\}$ states by $U$
and acts trivially on the other non-qutrit state has the following
matrix:
\begin{equation}
   \small
    \begin{pmatrix}
    \begin{tabular}{c|c|c|c}
    \begin{tabular}{c|c}
    \begin{tabular}{ccc}
    0 & 0 & 0\\
    0 & 0 & $\frac{i}{\sqrt{2}}$\\
    $\frac{1}{\sqrt{2}}$ & 0 & 0
    \end{tabular} & $\mathbf{0}^{\intercal}$\\ \hline
    \bf 0 & 1
    \end{tabular} &
    \begin{tabular}{c|c}
    \begin{tabular}{ccc}
    $\frac{1}{2}$ & 0 & 0\\
    $-\frac{i}{2}$ & 0 & 0\\
    0 & $\frac{i}{2}-\frac{1}{2\sqrt{2}}$ & $-\frac{i}{2\sqrt{2}}$
    \end{tabular} &$\mathbf{0}^{\intercal}$\\ \hline
    \bf 0 & 0
    \end{tabular} &
    \begin{tabular}{c|c}
    \begin{tabular}{ccc}
    0 & $\frac{i}{\sqrt{2}}$ & $-\frac{i}{2}$\\
    0 & 0 & $\frac{1}{2}$\\
    0 & 0 & 0
    \end{tabular} & $\mathbf{0}^{\intercal}$ \\ \hline
    \bf 0 & 0
    \end{tabular} &
    $\bf \hat{0}$\\ \hline
    \begin{tabular}{c|c}
    \begin{tabular}{ccc}
    $\frac{1}{2}$ & 0 & 0\\
    $\frac{i}{2}$ & 0 & 0\\
    0 & 1 & 0
    \end{tabular} & $\mathbf{0}^{\intercal}$ \\ \hline
    \bf 0 & 0
    \end{tabular} &
    \begin{tabular}{c|c}
    \begin{tabular}{ccc}
    0 & $\frac{1}{2}-\frac{i}{2\sqrt{2}}$ & $\frac{1}{2\sqrt{2}}$\\
    0 & $\frac{1}{2\sqrt{2}}$ & $-\frac{1}{2}-\frac{i}{2\sqrt{2}}$\\
    0 & 0 & 0
    \end{tabular} & $\mathbf{0}^{\intercal}$ \\ \hline
    \bf 0 & 1
    \end{tabular} &
    \begin{tabular}{c|c}
    \begin{tabular}{ccc}
    $\frac{1}{2}$ & 0 & 0\\
    $-\frac{i}{2}$ & 0 & 0\\
    0 & 0 & 0
    \end{tabular} &$\mathbf{0}^{\intercal}$\\ \hline
    \bf 0 & 0
    \end{tabular} &
    $\bf \hat{0}$\\ \hline
    \begin{tabular}{c|c}
    \begin{tabular}{ccc}
    0 & 0 & $\frac{1}{\sqrt{2}}$\\
    0 & 0 & 0\\
    0 & 0 & 0
    \end{tabular} &$\mathbf{0}^{\intercal}$\\ \hline
    \bf 0 & 0
    \end{tabular} &
    \begin{tabular}{c|c}
    \begin{tabular}{ccc}
    $\frac{1}{2}$ & 0 & 0\\
    $\frac{i}{2}$ & 0 & 0\\
    0 & $\frac{1}{2\sqrt{2}}$& $\frac{1}{2}-\frac{i}{2\sqrt{2}}$
    \end{tabular} &$\mathbf{0}^{\intercal}$\\ \hline
    \bf 0 & 0
    \end{tabular} &
    \begin{tabular}{c|c}
    \begin{tabular}{ccc}
    0 & 0 & $\frac{i}{2}$\\
    0 & $\frac{1}{\sqrt{2}}$ & $\frac{1}{2}$\\
    $-\frac{1}{\sqrt{2}}$ & 0 & 0
    \end{tabular} & $\mathbf{0}^{\intercal}$ \\ \hline
    \bf 0 & 1
    \end{tabular} &
    $\bf \hat{0}$\\ \hline
    $\bf \hat{0}$ & $\bf \hat{0}$ & $\bf \hat{0}$ & $\bf \hat{I}$
    \end{tabular}
   \end{pmatrix}\,,
     \label{Matrix_16}
\end{equation}
where $ \bf{\hat{0}} = \begin{pmatrix}
    0 & 0 & 0 & 0 \\
    0 & 0 & 0 & 0 \\
    0 & 0 & 0 & 0 \\
    0 & 0 & 0 & 0
\end{pmatrix}, \bf{\hat{I}} = \begin{pmatrix}
    1 & 0 & 0 & 0 \\
    0 & 1 & 0 & 0 \\
    0 & 0 & 1 & 0 \\
    0 & 0 & 0 & 1
\end{pmatrix}, \bf{0} = \begin{pmatrix}
    0 & 0 & 0
\end{pmatrix}.$

\section{Implementation of the Landau-Streater channel through the Werner-Holevo channel}\label{Sergey_matrix}
By definition, the WH channel\,\cite{WH} transforms the density matrix ($\rho \in \mathcal{B}(\mathcal{H}_d)$) as follows:
\begin{equation}
    \Phi_{\mathrm{WH}}[\rho] = \frac{1}{d-1}(\operatorname{Tr}[\rho]I-\rho^T)\,.
\end{equation}\label{WH}
After the unitary transformation
\begin{equation}
    W =
    \begin{pmatrix}
    0 & 0 & 1\\
    0 & -1 & 0\\
    1 & 0 & 0
    \end{pmatrix},
\end{equation}\label{T_matrix}
we get, accordingly, the LS channel:
\begin{equation}
    \Phi_{\mathrm{LS}}[\rho] = \Phi_{\mathrm{WH}}[W \rho W^{\dagger}].
\end{equation}
The WH channel can be represented as
\begin{equation}
    \Phi_{\mathrm{WH}}[\rho] = \operatorname{Tr}_{\mathrm{env}}(U(\rho_{\mathrm{env}}\otimes\rho)U^{\dagger}),
\end{equation}
where
\begin{equation}
\rho_{\mathrm{env}} = \ket{0}\bra{0},
\end{equation}\label{Sergey_env}
and
\begin{equation}
    U = \frac{1}{\sqrt{2}}
    \begin{pmatrix}
    \begin{tabular}{c|c|c}
    \begin{tabular}{ccc}
    0 & 1 & 0\\
    -1 & 0 & 0\\
    0 & 0 & 0
    \end{tabular} &
    \begin{tabular}{c}
    $U_{-1; 0}$
    \end{tabular} &
    \begin{tabular}{c}
    $U_{-1; 1}$
    \end{tabular}\\ \hline
    \begin{tabular}{ccc}
    0 & 0 & 1\\
    0 & 0 & 0\\
    -1 & 0 & 0
    \end{tabular} &
    \begin{tabular}{ccc}
    $U_{0; 0}$
    \end{tabular} &
    \begin{tabular}{c}
    $U_{0; 1}$
    \end{tabular}\\ \hline
    \begin{tabular}{ccc}
    0 & 0 & 0\\
    0 & 0 & 1\\
    0 & -1 & 0
    \end{tabular} &
    \begin{tabular}{ccc}
    $U_{1; 0}$
    \end{tabular} &
    \begin{tabular}{c}
    $U_{1; 1}$
    \end{tabular}
    \end{tabular}
    \end{pmatrix}.
\end{equation}
The only requirement for the $\{U_{m_j; 0}\}$ and $\{U_{m_j; 1}\}$ blocks is that $U$ is unitary.

With the selected method of encoding logical states of a qutrit by two qubits we implement the transformation $W$ as follows:
\begin{equation}
    \widetilde{W} =
    \begin{pmatrix}
    0 & 0 & 1 & 0\\
    0 & -1 & 0 & 0\\
    1 & 0 & 0 & 0\\
    0 & 0 & 0 & e^{i\varphi}
    \end{pmatrix}.
\end{equation}
\label{WHLS}
\begin{widetext}

This can be decomposed into elementary gates:
\begin{align}
W &=
\begin{pmatrix}
0 & 0 & 1 & 0\\
0 & -1 & 0 & 0\\
1 & 0 & 0 & 0\\
0 & 0 & 0 & -1
\end{pmatrix}
=
\begin{pmatrix}
1 & 0 & 0 & 0\\
0 & -1 & 0 & 0\\
0 & 0 & 1 & 0\\
0 & 0 & 0 & -1
\end{pmatrix}
\begin{pmatrix}
0 & 0 & 1 & 0\\
0 & 0 & 0 & 1\\
1 & 0 & 0 & 0\\
0 & 1 & 0 & 0
\end{pmatrix}
\begin{pmatrix}
1 & 0 & 0 & 0\\
0 & 0 & 0 & 1\\
0 & 0 & 1 & 0\\
0 & 1 & 0 & 0
\end{pmatrix}
\begin{pmatrix}
0 & 0 & 1 & 0\\
0 & 0 & 0 & 1\\
1 & 0 & 0 & 0\\
0 & 1 & 0 & 0
\end{pmatrix}
= \nonumber \\
&= Z_2 \times X_2\times \mathrm{CNOT}_{21} \times X_2 = i(Y_2\times \mathrm{CNOT}_{21} \times X_2),
\end{align}
\end{widetext}
where $X_k, Y_k, Z_k$ are Pauli gates acting on k-th qubit, and $CNOT_{kl}$ means CNOT gate where k-th qubit is a control qubit and l-th is the target qubit.

\begin{figure}[h]
\begin{minipage}[h]{0.47\linewidth}
\center{\includegraphics[width=1\linewidth]{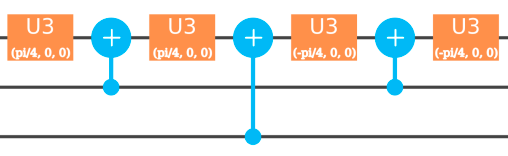}}
\end{minipage}
\hfill
\begin{minipage}[h]{0.47\linewidth}
\center{\includegraphics[width=1\linewidth]{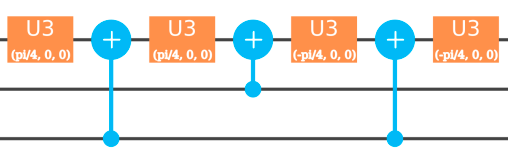}}
\end{minipage}
\vfill
\caption{Two different implementations of a quasi-Toffoli gate.}
\label{fig:quasitiffoli_options}
\end{figure}

For the effective decomposition into one-qubit gates, we replace the Toffoli gate by a quasi-Toffoli one.
It requires fewer CNOT gates (Fig.~\ref{fig:quasitiffoli_options}) and has the following matrix:
\begin{align}
\begin{pmatrix}
    1 & 0 & 0 & 0 & 0 & 0 & 0 & 0 \\
    0 & 1 & 0 & 0 & 0 & 0 & 0 & 0 \\
    0 & 0 & 1 & 0 & 0 & 0 & 0 & 0 \\
    0 & 0 & 0 & 1 & 0 & 0 & 0 & 0 \\
    0 & 0 & 0 & 0 & -1 & 0 & 0 & 0 \\
    0 & 0 & 0 & 0 & 0 & 1 & 0 & 0 \\
    0 & 0 & 0 & 0 & 0 & 0 & 0 & 1 \\
    0 & 0 & 0 & 0 & 0 & 0 & 1 & 0 \\
\end{pmatrix}.
\end{align}

\begin{figure}[h]
\begin{center}
\includegraphics[scale=0.3]{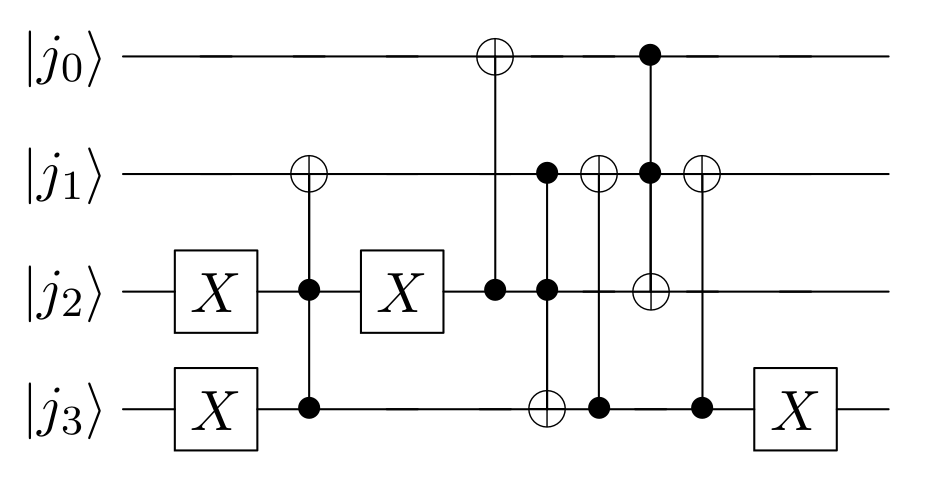}
\caption{Algorithm for line permutations. Qubits $\ket{j_2}$ and $\ket{j_3}$ encode the environment qutrit
and qubits $\ket{j_0}$ and $\ket{j_1}$ encode the channel-transformed spin-1 particle.}
\label{fig:String_swap}
\end{center}
\end{figure}

In the matrix emulation, see Sec.\,\ref{Sergey_matrix},  by qubits that is described in Sec.\,\ref{qutrit_emulation}, the
impact on the unused state (which does not encode a logical qutrit) could be any if it preserves the unitarity of the matrix.
Taking advantage of this condition, we represent the matrix for Werner-Holevo channel as:
\begin{equation}
    \widetilde{U} =
    \frac{1}{\sqrt{2}}
    \begin{pmatrix}
    \begin{tabular}{c|c}
         \begin{tabular}{ccc}
         0 & 1 & 0\\
         -1 & 0 & 0\\
         0 & 0 & 0\\
         0 & 0 & 0\\ \hline
         0 & 0 & 1\\
         0 & 0 & 0\\
         -1 & 0 & 0\\
         0 & 0 & 0\\ \hline
         0 & 0 & 0\\
         0 & 0 & 1\\
         0 & -1 & 0\\
         0 & 0 & 0\\ \hline
         0 & 0 & 0\\
         0 & 0 & 0\\
         0 & 0 & 0\\
         0 & 0 & 0
         \end{tabular}
         & $\widetilde{U}'$
    \end{tabular}
    \end{pmatrix}.
\end{equation}
The columns of $\widetilde{U}'$ are arbitrary except that the $\widetilde{U}$ matrix must satisfy the unitary constraint.
Using the quasi-Toffoli and CNOT gates, we make lines permutations, $S$, (Fig.~\ref{fig:String_swap}) of the $\widetilde{U}$
matrix such that its absolute values are equal to a tensor product of one-qubit gates.

As a result, we get the following matrix:
\begin{align}
S\widetilde{U} =
\frac{1}{\sqrt{2}}
\begin{pmatrix}
\begin{tabular}{c|c}
     \begin{tabular}{ccc}
     0 & 0 & 0\\
     0 & 0 & 1\\
     0 & 1 & 0\\
     -1 & 0 & 0\\ \hline
     0 & 0 & 0\\
     0 & 0 & 0\\
     0 & 0 & 0\\
     0 & 0 & 0\\ \hline
     0 & 0 & 0\\
     0 & 0 & 1\\
     0 & -1 & 0\\
     -1 & 0 & 0\\ \hline
     0 & 0 & 0\\
     0 & 0 & 0\\
     0 & 0 & 0\\
     0 & 0 & 0
     \end{tabular}
     & $S\widetilde{U}'$
\end{tabular}
\end{pmatrix},
\end{align}
which is almost equal (some elements might have different phases because we used the quasi-Toffoli gate instead the Toffoli one) to:
\begin{center}
$\frac{1}{\sqrt{2}}
\begin{pmatrix}
\begin{tabular}{c|c|c|c}
     \begin{tabular}{cc}
     \bf 0 & $\bf\hat\sigma_x$\\
     $\bf\hat\sigma_x$ & \bf 0
     \end{tabular} &
     \begin{tabular}{cc}
     \bf 0 & \bf 0\\
     \bf 0 & \bf 0
     \end{tabular} &
     \begin{tabular}{cc}
     \bf 0 & $\bf\hat\sigma_x$\\
     $\bf\hat\sigma_x$ & \bf 0
     \end{tabular} &
     \begin{tabular}{cc}
     \bf 0 & \bf 0\\
     \bf 0 & \bf 0
     \end{tabular}\\ \hline
     \begin{tabular}{cc}
     \bf 0 & \bf 0\\
     \bf 0 & \bf 0
     \end{tabular} &
     \begin{tabular}{cc}
     \bf 0 & $\bf\hat\sigma_x$\\
     $\bf\hat\sigma_x$ & \bf 0
     \end{tabular} &
     \begin{tabular}{cc}
     \bf 0 & \bf 0\\
     \bf 0 & \bf 0
     \end{tabular} &
     \begin{tabular}{cc}
     \bf 0 & $\bf\hat\sigma_x$\\
     $\bf\hat\sigma_x$ & \bf 0
     \end{tabular}\\ \hline
     \begin{tabular}{cc}
     \bf 0 & $\bf\hat\sigma_x$\\
     $\bf\hat\sigma_x$ & \bf 0
     \end{tabular} &
     \begin{tabular}{cc}
     \bf 0 & \bf 0\\
     \bf 0 & \bf 0
     \end{tabular} &
     \begin{tabular}{cc}
     \bf 0 & $\bf{-\hat\sigma_x}$\\
     $\bf{-\hat\sigma_x}$ & \bf 0
     \end{tabular} &
     \begin{tabular}{cc}
     \bf 0 & \bf 0\\
     \bf 0 & \bf 0
     \end{tabular}\\ \hline
     \begin{tabular}{cc}
     \bf 0 & \bf 0\\
     \bf 0 & \bf 0
     \end{tabular} &
     \begin{tabular}{cc}
     \bf 0 & $\bf\hat\sigma_x$\\
     $\bf\hat\sigma_x$ & \bf 0
     \end{tabular} &
     \begin{tabular}{cc}
     \bf 0 & \bf 0\\
     \bf 0 & \bf 0
     \end{tabular} &
     \begin{tabular}{cc}
     \bf 0 & $\bf{-\hat\sigma_x}$\\
     $\bf{-\hat\sigma_x}$ & \bf 0
     \end{tabular}\\
\end{tabular}
\end{pmatrix}=$
\end{center}

\begin{align}
    &=\bf{H}\otimes\bf{\bf{Id}}\otimes\bf{\hat\sigma_x}\otimes{\bf\hat\sigma_x} = \nonumber \\ &=
    \bf\begin{pmatrix}
    \frac{1}{\sqrt{2}} & \frac{1}{\sqrt{2}}\\
    \frac{1}{\sqrt{2}} & -\frac{1}{\sqrt{2}}
    \end{pmatrix}\otimes
    \bf\begin{pmatrix}
    1 & 0\\
    0 & 1
    \end{pmatrix}\otimes
    \bf\begin{pmatrix}
    0 & 1\\
    1 & 0
    \end{pmatrix}\otimes
    \bf\begin{pmatrix}
    0 & 1\\
    1 & 0
    \end{pmatrix}.
\end{align}

\begin{figure}[h]
\begin{minipage}[h]{0.94\linewidth}
\center{\includegraphics[width=1\linewidth]{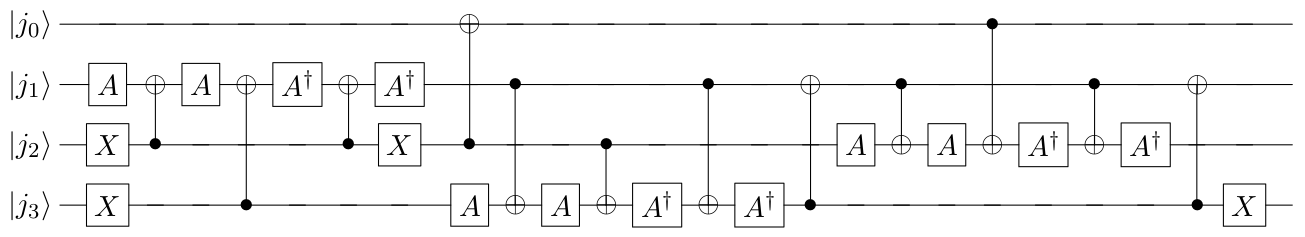}} $S_1$
\end{minipage}
\vfill
\begin{minipage}[h]{0.94\linewidth}
\center{\includegraphics[width=1\linewidth]{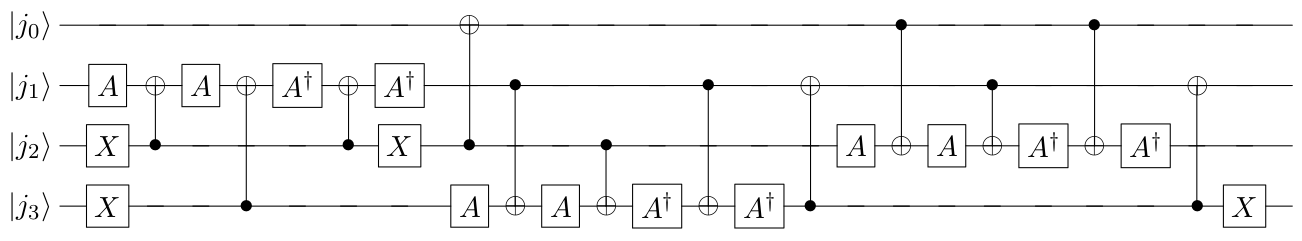}} $S_2$
\end{minipage}
\vfill
\begin{minipage}[h]{0.94\linewidth}
\center{\includegraphics[width=1\linewidth]{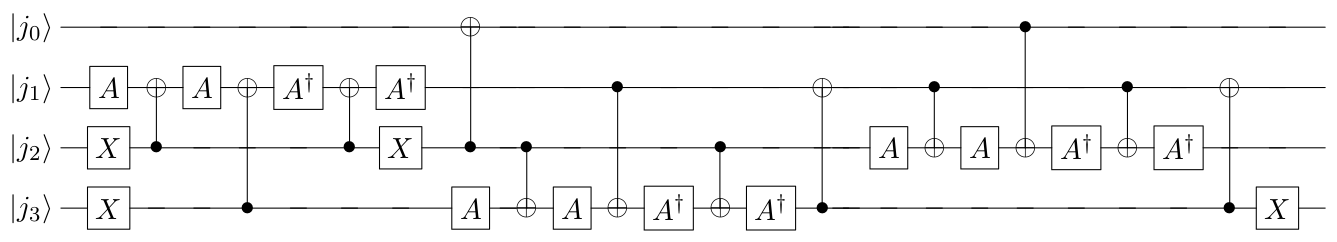}} $S_3$
\end{minipage}
\vfill
\begin{minipage}[h]{0.94\linewidth}
\center{\includegraphics[width=1\linewidth]{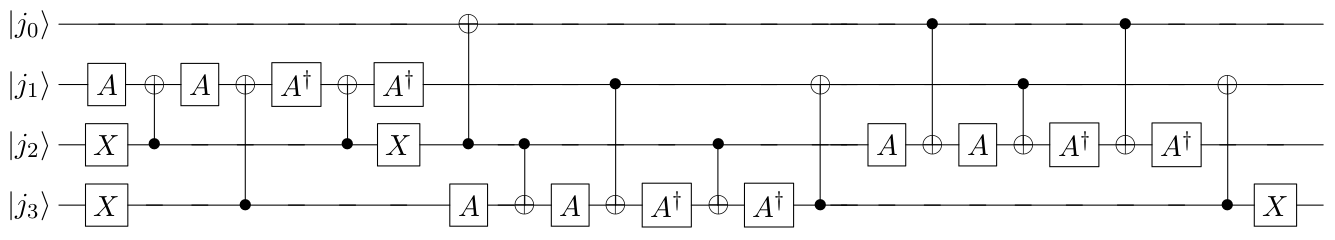}} $S_4$
\end{minipage}
\caption{Implementations of the algorithm shown in Fig.~\ref{fig:String_swap}, using the CNOT decomposition of quasi-Toffoli gates.
Here $A = U_3(\pi/4, 0, 0) = \begin{pmatrix}
    \cos(\frac{\pi}{8}) & -\sin(\frac{\pi}{8}) \\
    \sin(\frac{\pi}{8}) & \cos(\frac{\pi}{8})
\end{pmatrix}$.}
\label{fig:tensor_product}
\end{figure}

A decomposition into a tensor product of simple gates exist only in 4  out of 8 cases (Fig.~\ref{fig:tensor_product}):
\begin{equation}
    S_1 \widetilde{U}=
    \bf\begin{pmatrix}
    \frac{1}{\sqrt{2}} & \frac{1}{\sqrt{2}}\\
    -\frac{1}{\sqrt{2}} & \frac{1}{\sqrt{2}}
    \end{pmatrix}\otimes
    \bf\begin{pmatrix}
    1 & 0\\
    0 & 1
    \end{pmatrix}\otimes
    \bf\begin{pmatrix}
    0 & 1\\
    1 & 0
    \end{pmatrix}\otimes
    \bf\begin{pmatrix}
    0 & 1\\
    -1 & 0
    \end{pmatrix},
\end{equation}

\begin{equation}
    S_2 \widetilde{U}=
    \bf\begin{pmatrix}
    \frac{1}{\sqrt{2}} & \frac{1}{\sqrt{2}}\\
    -\frac{1}{\sqrt{2}} & \frac{1}{\sqrt{2}}
    \end{pmatrix}\otimes
    \bf\begin{pmatrix}
    1 & 0\\
    0 & 1
    \end{pmatrix}\otimes
    \bf\begin{pmatrix}
    0 & 1\\
    1 & 0
    \end{pmatrix}\otimes
    \bf\begin{pmatrix}
    0 & 1\\
    -1 & 0
    \end{pmatrix},
\end{equation}

\begin{equation}
    S_3 \widetilde{U}=
    \bf\begin{pmatrix}
    \frac{1}{\sqrt{2}} & \frac{1}{\sqrt{2}}\\
    \frac{1}{\sqrt{2}} & -\frac{1}{\sqrt{2}}
    \end{pmatrix}\otimes
    \bf\begin{pmatrix}
    1 & 0\\
    0 & 1
    \end{pmatrix}\otimes
    \bf\begin{pmatrix}
    0 & 1\\
    1 & 0
    \end{pmatrix}\otimes
    \bf\begin{pmatrix}
    0 & -1\\
    1 & 0
    \end{pmatrix},
\end{equation}

\begin{equation}
    S_4 \widetilde{U}=
    \bf\begin{pmatrix}
    \frac{1}{\sqrt{2}} & \frac{1}{\sqrt{2}}\\
    \frac{1}{\sqrt{2}} & -\frac{1}{\sqrt{2}}
    \end{pmatrix}\otimes
    \bf\begin{pmatrix}
    1 & 0\\
    0 & 1
    \end{pmatrix}\otimes
    \bf\begin{pmatrix}
    0 & 1\\
    1 & 0
    \end{pmatrix}\otimes
    \bf\begin{pmatrix}
    0 & -1\\
    1 & 0
    \end{pmatrix}.
\end{equation}

\begin{figure}[h]
\begin{minipage}[h]{0.47\linewidth}
\center{\includegraphics[width=1\linewidth]{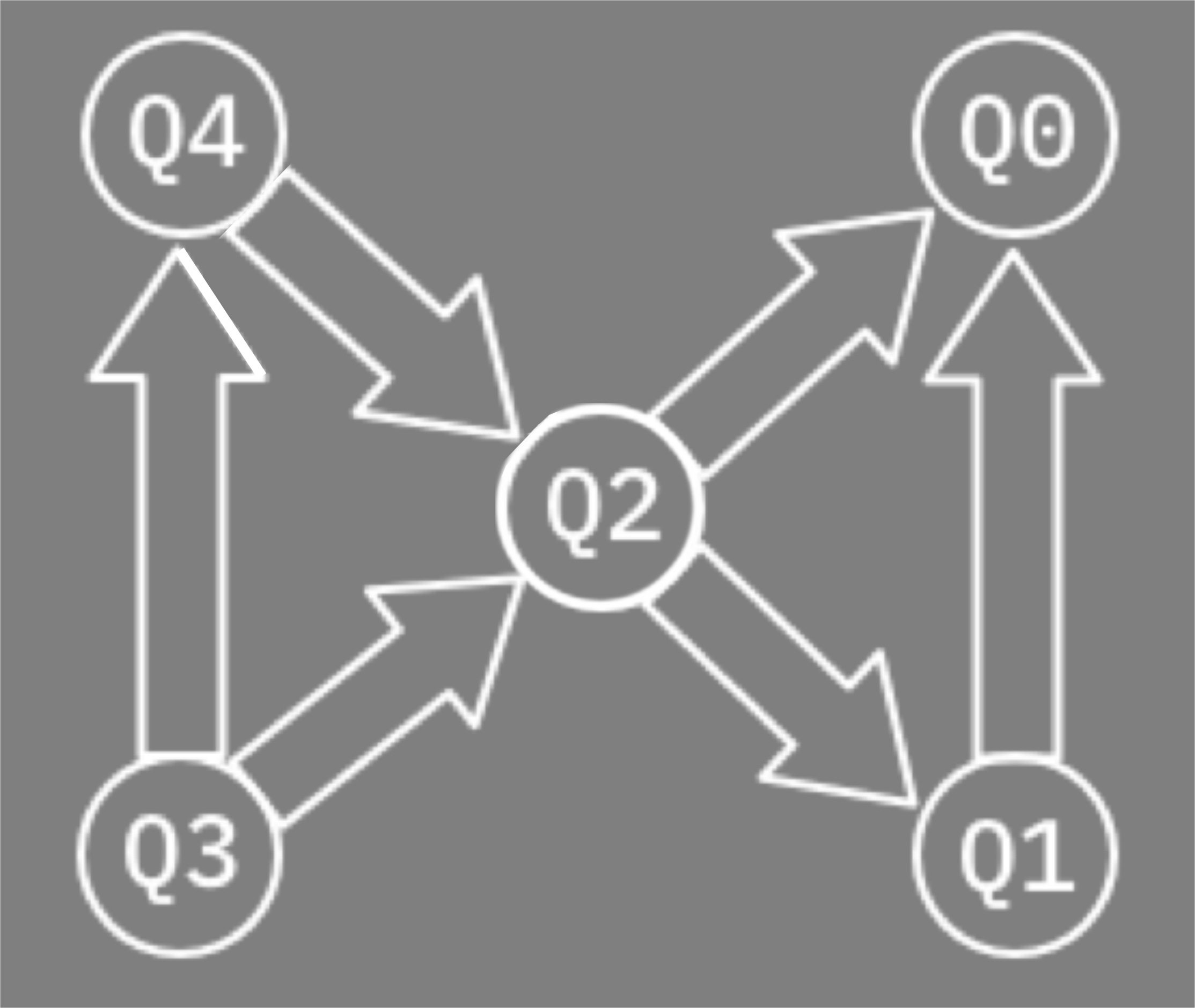}
\caption{Coupling map of two-qubit gates on a 5-qubit machine}\label{fig:arch_ibmqx4}}
\end{minipage}
\hfill
\begin{minipage}[h]{0.47\linewidth}
\center{\includegraphics[width=1\linewidth]{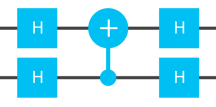}\caption{Swap algorithm for the target and control qubits of a CNOT gate.}
\label{fig:CNOT_redirect}}
\end{minipage}
\end{figure}

\begin{figure}[h]
        \center{\includegraphics[width=0.8\linewidth]{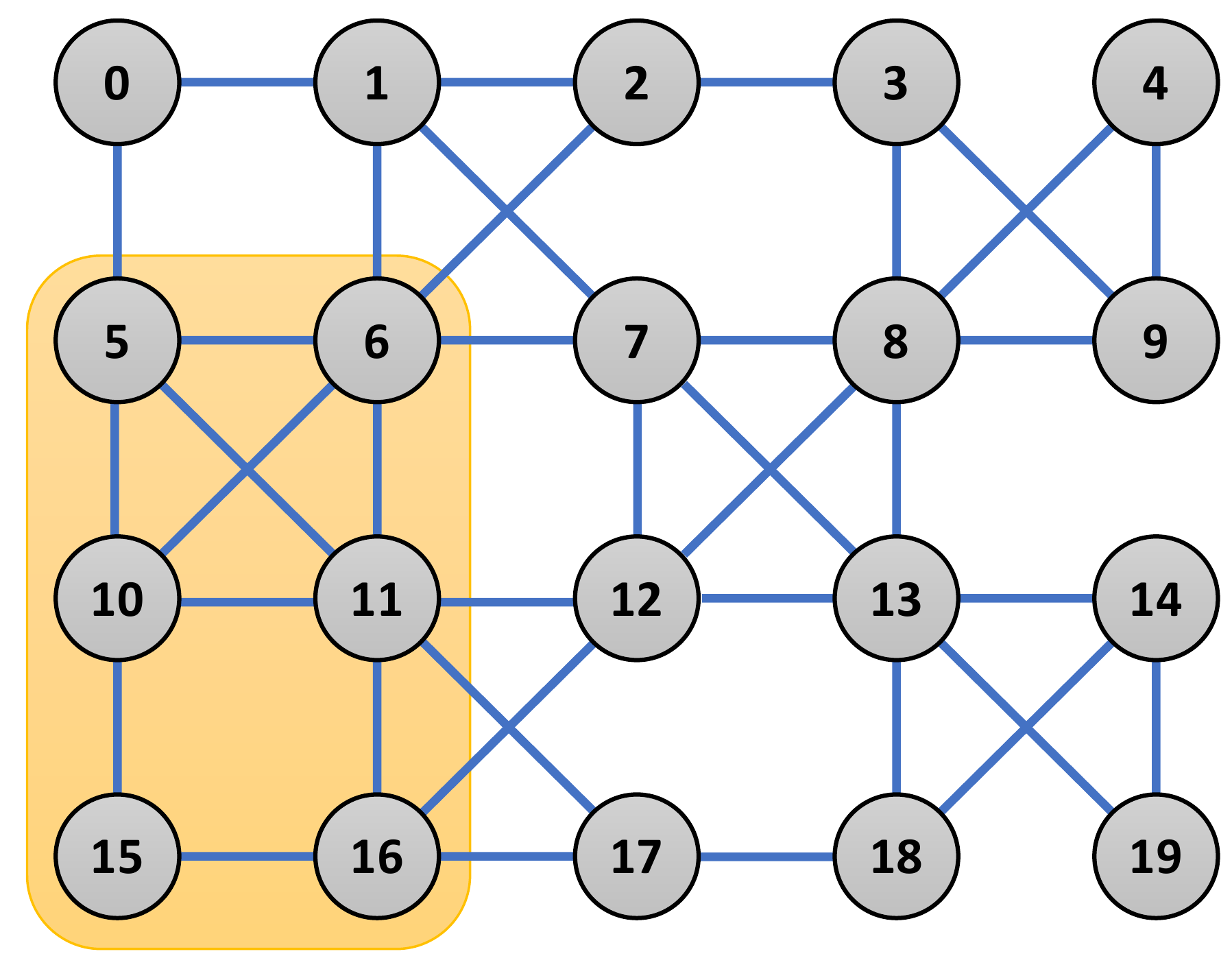}}
            \caption{Connectivity map of IBM's 20-qubit machine \textsf{tokyo}. The (yellow) highlighted qubits where used to implement the quantum channels.}
            \label{fig:conmap_tokyo}
\end{figure}

Not every CNOT gate can be performed on IBM quantum computers due to missing connections, see connectivity map for the 5-qubit machine in Fig.~\ref{fig:arch_ibmqx4} and 20-qubit machine in Fig.~\ref{fig:conmap_tokyo}.

On the 5-qubit machine each couple of qubits in the coupling map has only one target and control qubit.
Therefore, if we need to swap the control and target qubit,
we need to use the method shown in Fig.~\ref{fig:CNOT_redirect} adding Hadamard gates.
We choose a configuration in the Fig. ~\ref{fig:String_swap}
such that the number of two-qubit CNOT gates is minimal because CNOT gate has the greatest error.

\begin{figure*}[!ht]
\begin{center}
\includegraphics[width = \linewidth]{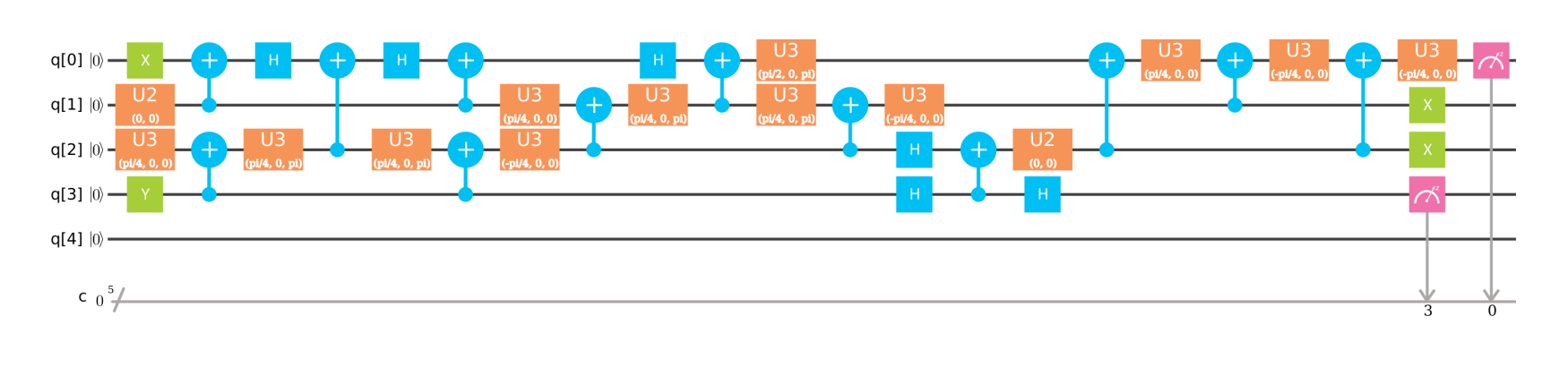}
\caption{Implementation of the $S_4$ configuration (Fig.~\ref{fig:String_swap}) on a 5-quibit machine regarding its coupling map.}
\label{fig:WH}
\end{center}
\end{figure*}

On this machine, we use the $S_4$ configuration implementing the
Werner-Holevo channel. The sequence of gates for this is shown in
figure~\ref{fig:WH} ($U1(\lambda)$, $U2(\varphi, \lambda)$, and
$U3(\theta, \varphi, \lambda)$  are a full set of one-qubit gates
on IBM machines). Qubits $\ket{q3, q0}$ encode $\ket{j0, j1}$
(channel qubits) and $\ket{q2, q1}$ represent $\ket{j2, j3}$
(environment qubits).

The $\widetilde{W}$ transformation requires a CNOT gate between the system qubits ($\ket{q3, q0}$).
However, the coupling map of the 5-qubit machine does not allow to place a CNOT operator there.
Therefore, we have to use three CNOT gates instead of a single one and place them in accordance with the coupling map (Fig.~\ref{fig:arch_ibmqx4}).

\begin{figure*}[!ht]
\begin{center}
\includegraphics[width = \linewidth]{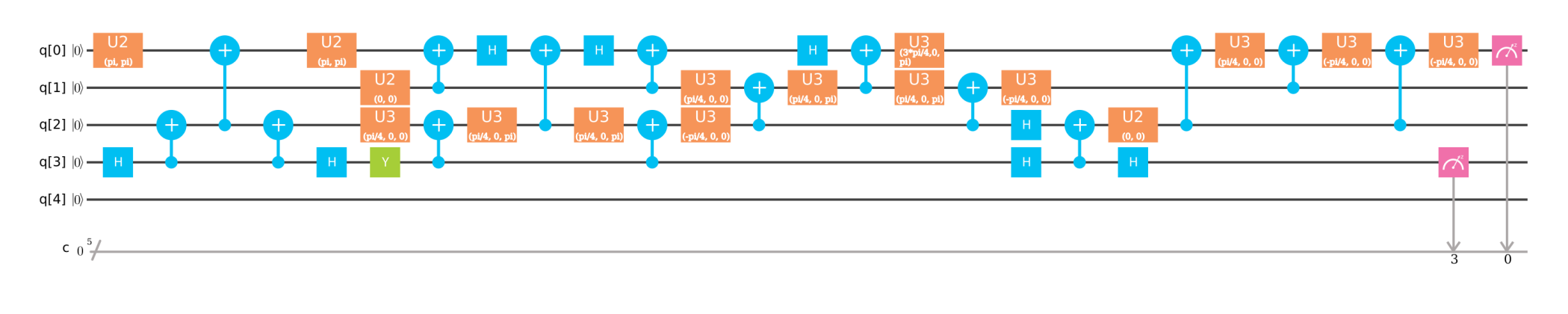}
\caption{Realization of Landau-Streater channel based on the Werner-Holevo channel.}
\label{fig:WHLS_pic}
\end{center}
\end{figure*}

Finally, we realize the unitary interaction U (formula
\ref{Ksenia_matrix}) with the use of $S_4 \widetilde{U}$ (Fig.
\ref{fig:WH}) and $\widetilde{W}$ transformations (Fig.
\ref{fig:WHLS_pic}).

\section{Measurements}
In order to predict the final quantum state for an arbitrary initial state, we performed the transformation of each basis density matrix per channel.
Explicitly, we use the following 9 matrices as inputs:
\begin{equation}
\scalebox{0.87}{
$\begin{split}
     \rho_1 &= \begin{pmatrix}
        1 & 0 & 0\\
        0 & 0 & 0\\
        0 & 0 & 0
        \end{pmatrix},
    & \rho_2 &= \begin{pmatrix}
        0 & 0 & 0\\
        0 & 1 & 0\\
        0 & 0 & 0
        \end{pmatrix},
    & \rho_3 &= \begin{pmatrix}
        0 & 0 & 0\\
        0 & 0 & 0\\
        0 & 0 & 1
        \end{pmatrix}, \\
    \rho_4 &= \frac{1}{2}
        \begin{pmatrix}
        1 & 1 & 0\\
        1 & 1 & 0\\
        0 & 0 & 0
        \end{pmatrix},
    & \rho_5 &= \frac{1}{2}
        \begin{pmatrix}
        1 & 0 & 1\\
        0 & 0 & 0\\
        1 & 0 & 1
        \end{pmatrix},
    & \rho_6 &= \frac{1}{2}
        \begin{pmatrix}
        0 & 0 & 0\\
        0 & 1 & 1\\
        0 & 1 & 1
        \end{pmatrix}, \\
    \rho_7 &= \frac{1}{2}
        \begin{pmatrix}
        1 & -i & 0\\
        i & 1 & 0\\
        0 & 0 & 0
        \end{pmatrix},
    & \rho_8 & = \frac{1}{2}
        \begin{pmatrix}
        1 & 0 & -i\\
        0 & 0 & 0\\
        i & 0 & 1
    \end{pmatrix},
    & \rho_9 &= \frac{1}{2}
        \begin{pmatrix}
        0 & 0 & 0\\
        0 & 1 & -i\\
        0 & i & 1
        \end{pmatrix}
    \label{eq:bases_denstity_matrices}
    \end{split}
$}
\end{equation}

\begin{figure}[!h]
\begin{minipage}[h]{0.22\linewidth}
\center{\includegraphics[width=1\linewidth]{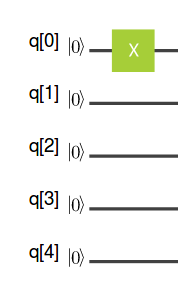}} $\ket{\psi_2} = \ket{01}$
\end{minipage}
\hfill
\begin{minipage}[h]{0.22\linewidth}
\center{\includegraphics[width=1\linewidth]{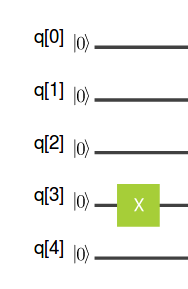}} $\ket{\psi_3} = \ket{10}$
\end{minipage}
\hfill
\begin{minipage}[h]{0.22\linewidth}
\center{\includegraphics[width=1\linewidth]{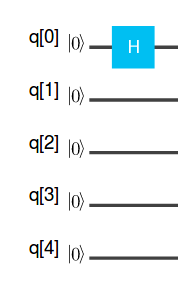}} $\ket{\psi_4} = \frac{\ket{00}+\ket{01}}{\sqrt{2}}$
\end{minipage}\vfill
\begin{minipage}[h]{0.22\linewidth}
\center{\includegraphics[width=1\linewidth]{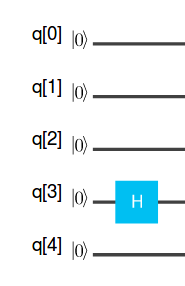}} $\ket{\psi_5} = \frac{\ket{00}+\ket{10}}{\sqrt{2}}$
\end{minipage}
\hfill
\begin{minipage}[h]{0.47\linewidth}
\center{\includegraphics[width=1\linewidth]{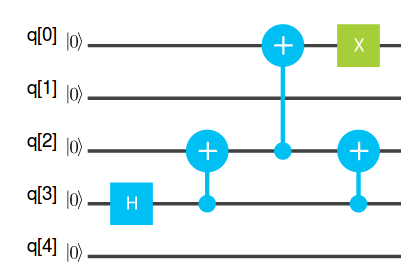}} $\ket{\psi_6} = \frac{\ket{01}+\ket{10}}{\sqrt{2}}$
\end{minipage}
\hfill
\begin{minipage}[h]{0.22\linewidth}
\center{\includegraphics[width=1\linewidth]{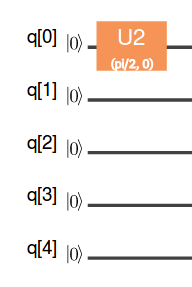}} $\ket{\psi_7} = \frac{\ket{00}+i\ket{01}}{\sqrt{2}}$
\end{minipage}
\vfill
\begin{minipage}[h]{0.22\linewidth}
\center{\includegraphics[width=1\linewidth]{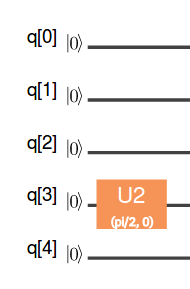}} $\ket{\psi_8} = \frac{\ket{00}+i\ket{10}}{\sqrt{2}}$
\end{minipage}
\hfill
\begin{minipage}[h]{0.47\linewidth}
\center{\includegraphics[width=1\linewidth]{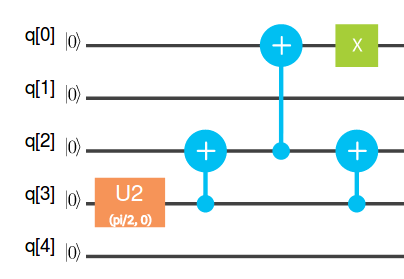}} $\ket{\psi_9} = \frac{\ket{01}+i\ket{10}}{\sqrt{2}}$
\end{minipage}
\caption{Preparation of initial states. If all qubits were connected to each other on the 5-quibit machine, only one CNOT gate would be required for
$\ket{\psi_6}$ and $\ket{\psi_9}$, but given the current coupling map, we need to use 3 CNOTs.}
\label{fig:state_prep}
\end{figure}

After the transformation by the channel, we perform a tomography~\cite{Tomo} of the system qutrit.
Knowing these density matrices, we can reconstruct a channel using its linear property:
\begin{align}
    \Phi\left(\sum_i a_i \rho_i\right) = \sum_i a_i \Phi(\rho_i).
\end{align}
Any of basis states can be prepared using unitary transformations, see Fig.~\ref{fig:state_prep} (where $\rho_k = \ket{\psi_k} \bra{\psi_k}$).

In practice, IBM has implemented the  tomography algorithm in their python library
(\href{https://github.com/QISKit}{QISKit}\cite{qiskit}).
We use it instead of our own realization.

As a quality test, we use the following expression to calculate the fidelity~\cite{fidelity} between the theoretical and experimental output density matrix for same inputs:
\begin{equation}
    F(\sigma_1, \sigma_2) = \left(\operatorname{Tr}\sqrt{\sqrt{\sigma_1}\sigma_2\sqrt{\sigma_1}}\right)^2\,,
\end{equation}
where $F(\sigma_1, \sigma_2)=F(\sigma_2, \sigma_1)$.

The \href{https://github.com/QISKit}{QISKit}\cite{qiskit} toolkit also includes a simulator of an ideal quantum computer, which has no measurement or environment coupling errors. We thus obtain the theoretically expected results.
Results of all tomography experiments on simulator for the WH $\Phi_{\mathrm{WH}}[\rho_i]$ and LS  $\Phi_{\mathrm{LS}}[\rho_i]$
channels fit well to the theoretical expectations (fidelity of both channels is nearly 0.99).

\section{Choi matrix of the Landau-Streater channel and comparison with theoretical expectations}
Any quantum channel acting on the Hilbert space $\mathcal{H}_N$  is connected to a linear map in $\mathcal{H}_N \otimes \mathcal{H}_N$ (Choi–Jamiołkowski isomorphism).
The duality between channels and states refers to the map:
\begin{align}
    \mathrm{\Phi} \rightarrow \Omega = (Id \otimes \Phi) (\ket{\psi_+} \bra{\psi_+}),
    \label{eq:direct_choi}
\end{align}
where $\mathrm{\Phi}$ is the quantum channel, $Id$ the identity channel ($\forall \rho \in \mathcal{B}(\mathcal{H}_N): Id(\rho) = \rho$), and
$ \ket{\psi_+} = \frac{1}{\sqrt{N}} \sum_{i=1}^N \ket{i} \ket{i} $.
\begin{align}
    \Omega &= (Id \otimes \Phi) (\ket{\psi_+} \bra{\psi_+}) = \nonumber \\ &= \frac{1}{N} (Id \otimes \Phi)
    (\sum_{i=1}^N \sum_{k=1}^N \ket{i} \ket{i} \bra{k} \bra{k}) = \nonumber \\ &= \frac{1}{N} (Id \otimes \Phi)
    (\sum_{i=1}^N \sum_{k=1}^N \ket{i} \bra{k} \otimes \ket{i} \bra{k}) = \nonumber \\ &= \frac{1}{N} \sum_{i=1}^N \sum_{k=1}^N
    \ket{i} \bra{k} \otimes \mathrm{\Phi} (\ket{i} \bra{k}).
\end{align}
This formula allows one to restore the Choi matrix by measuring final states.
As one sees, $\ket{i}\bra{k}$ products might not be the density operators ($\operatorname{Tr} \ket{i}\bra{k} = 0$, if $ i \neq k$).
But each term $\ket{i} \bra{k}$ can be represented as a linear combination of physical matrices.
For convenience, $ E_{i,j}$ denotes $ \ket{i} \bra{j} $.
In case the dimension is equal to 3, we get following expression:
\begin{align}
    E_{i,j} &=
    \begin{pmatrix}
        \ket{0} \bra{0} \\
        \ket{0} \bra{1} \\
        \ket{0} \bra{2} \\
        \ket{1} \bra{0} \\
        \ket{1} \bra{1} \\
        \ket{1} \bra{2} \\
        \ket{2} \bra{0} \\
        \ket{2} \bra{1} \\
        \ket{2} \bra{2} \\
    \end{pmatrix} = \sum_{k=1}^9 a_{i,j}^k \cdot \Re_k = \nonumber \\ &=
    \begin{pmatrix}
        1 & 0 & 0 & 0 & 0 & 0 & 0 & 0 & 0 \\
        -\frac{1 + i}{2} & -\frac{1 + i}{2} & 0 & 1 & 0 & 0 & i & 0 & 0 \\
        -\frac{1 + i}{2} & 0 & -\frac{1 + i}{2} & 0 & 1 & 0 & 0 & i & 0 \\
        -\frac{1 - i}{2} & -\frac{1 - i}{2} & 0 & 1 & 0 & 0 & -i & 0 & 0 \\
        0 & 1 & 0 & 0 & 0 & 0 & 0 & 0 & 0 \\
        0 & -\frac{1 + i}{2} & -\frac{1 + i}{2} & 0 & 0 & 1 & 0 & 0 & i \\
        -\frac{1 - i}{2} & 0 & -\frac{1 - i}{2} & 0 & 1 & 0 & 0 & -i & 0 \\
        0 & -\frac{1 - i}{2} & -\frac{1 - i}{2} & 0 & 0 & 1 & 0 & 0 & -i \\
        0 & 0 & 1 & 0 & 0 & 0 & 0 & 0 & 0 \\
    \end{pmatrix} \cdot \Re_k,
\end{align}
where $\Re_k =
\begin{pmatrix}
    \ket{0} \bra{0} \\
    \ket{1} \bra{1} \\
    \ket{2} \bra{2} \\
    \frac{1}{2} ( \ket{0} + \ket{1} ) ( \bra{0} + \bra{1} ) \\
    \frac{1}{2} ( \ket{0} + \ket{2} ) ( \bra{0} + \bra{2} ) \\
    \frac{1}{2} ( \ket{1} + \ket{2} ) ( \bra{1} + \bra{2} ) \\
    \frac{1}{2} ( \ket{0} + i \ket{1} ) ( \bra{0} - i \bra{1} ) \\
    \frac{1}{2} ( \ket{0} + i \ket{2} ) ( \bra{0} - i \bra{2} ) \\
    \frac{1}{2} ( \ket{1} + i \ket{2} ) ( \bra{1} - i \bra{2} ) \\
\end{pmatrix}
$.
Then, the Choi matrix $\Omega$ can be expressed as:
\begin{align}
    \label{eq:minimum_qubit_choi}
    \Omega &= \sum_{i=1}^N \sum_{j=1}^N E_{i,j} \otimes \Phi(E_{i,j}) = \nonumber \\
           &= \sum_{i=1}^N \sum_{j=1}^N E_{i,j}
    \otimes \sum_{k=1}^{N^2} a_{i,j}^k \Phi(\Re_k).
\end{align}
For illustrative purposes, we present it as a block matrix:
\begin{align}
    \label{eq:minimum_qubit_choi_block}
    \Omega =
    \begin{pmatrix}
        \sum_{k=1}^{N^2} a_{0,0}^k \Phi(\Re_k) & \sum_{k=1}^{N^2} a_{0,1}^k \Phi(\Re_k) & \sum_{k=1}^{N^2} a_{0,2}^k \Phi(\Re_k) \\
        \sum_{k=1}^{N^2} a_{1,0}^k \Phi(\Re_k) & \sum_{k=1}^{N^2} a_{1,1}^k \Phi(\Re_k) & \sum_{k=1}^{N^2} a_{1,2}^k \Phi(\Re_k) \\
        \sum_{k=1}^{N^2} a_{2,0}^k \Phi(\Re_k) & \sum_{k=1}^{N^2} a_{2,1}^k \Phi(\Re_k) & \sum_{k=1}^{N^2} a_{2,2}^k \Phi(\Re_k) \\
    \end{pmatrix}
\end{align}
Consequently, in order to build the Choi matrix $\Omega$, it is sufficient to know how the channel acts on the basis matrices $\Re$.

The Choi state ~\eqref{eq:direct_choi} cannot be physically realized on IBM a 5-qubit machine because needs an extra qutrit (plus 2 qubits and 6 qubits in total).
The connectivity map of 15-qubit machines also does not allow a direct implementation.
Only on a 20-qubit computer (see connectiivty map in Fig~\ref{fig:conmap_tokyo}) we can built the Choi matrix using formula \eqref{eq:direct_choi}.
In order to be able to utilize the 5-qubit machine, we used another expression, Eqs.~\eqref{eq:minimum_qubit_choi} \& \eqref{eq:minimum_qubit_choi_block} to reconstruct the Choi matrix $\Omega$.

\begin{figure}[!htbp]
\begin{center}
     \begingroup
     \fontsize{14pt}{16pt}\selectfont
         \textbf{5-qubit computer(ibmqx4)}
     \endgroup
    \begin{minipage}[t]{\linewidth}
        \includegraphics[width = 0.94\linewidth]{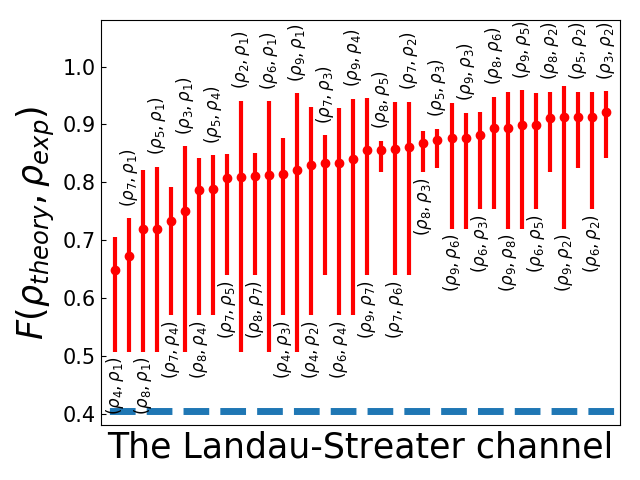}
    \end{minipage}
    \\
    \begin{minipage}[t]{\linewidth}
        \includegraphics[width = 0.94\linewidth]{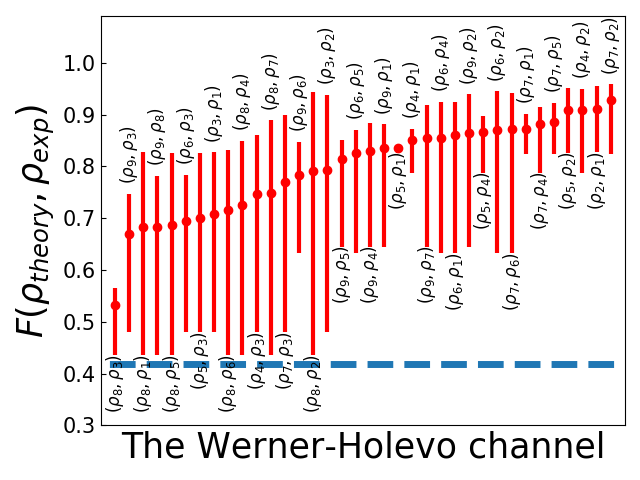}
    \end{minipage}
\caption{
    Fidelity by pairs of density matrices. Every pair stick contains maximum, minimum, and average (a point on a stick)
    The dashed line represents the fidelity between theoretical and experimental Choi matrices $\Omega$.
}
\label{fig:more_fidelity}
\end{center}
\end{figure}

Using this procedure, we get the Choi matrices $\Omega$ for the Landau-Streater and Werner-Holevo channels.
We compared them with the theoretical expectations, and obtained the values 0.406 and 0.419  for the fidelity ($F(\Omega_{\mathrm{theory}}, \Omega_{\mathrm{exp}}) = (\operatorname{Tr} \sqrt{\sqrt{\Omega_{\mathrm{theory}}} \Omega_{\mathrm{exp}} \sqrt{\Omega_{\mathrm{theory}}}})^2$), respectively.

Since the Choi matrix $\Omega$ contains all information about the channel map, we can reconstruct it according to the following expression:
\begin{align}
    \Phi(\rho) = \mathrm{Tr}_1 \big( (\rho^{\intercal} \otimes Id) \cdot \Omega \big)\,.
    \label{eq:channel_from_choi}
\end{align}

In figure~\ref{fig:more_fidelity}, we show the fidelity between theoretical and experimental results for both channels.
For each pair of the basis density matrices, we calculated the fidelity of
$ \lambda \rho_a + (1-\lambda) \rho_b \text{ for } \forall \lambda \in [0, 1] $ using Eq.~\eqref{eq:channel_from_choi}.
Each bar for the pairs  shows maximum, minimum, and average (point on the bar).

\begin{figure}[!ht]
\begin{center}
\includegraphics[width = 0.9 \linewidth]{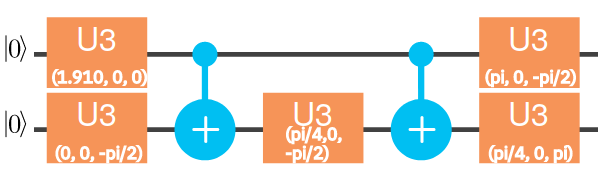}
\caption{Preparation of a qutrit superposition state $\frac{\ket{00} + \ket{01} + \ket{10}}{\sqrt{3}}$.
1.910 is the approximate value of $2 \arccos{\frac{1}{\sqrt{3}}}$}
\label{fig:superposition}
\end{center}
\end{figure}

\begin{figure}[!htbp]
\begin{center}
     \begingroup
     \fontsize{14pt}{16pt}\selectfont
         \textbf{20-qubit computer(tokyo)}
     \endgroup
    \begin{minipage}[t]{\linewidth}
        \includegraphics[width = 0.94\linewidth]{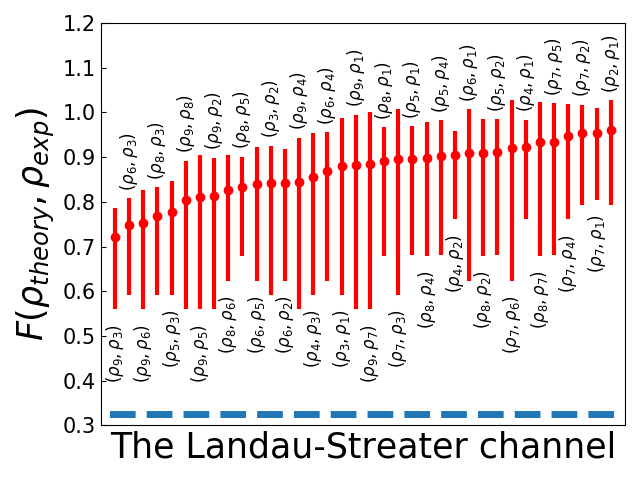}
    \end{minipage}
    \\
    \begin{minipage}[t]{\linewidth}
        \includegraphics[width = 0.94\linewidth]{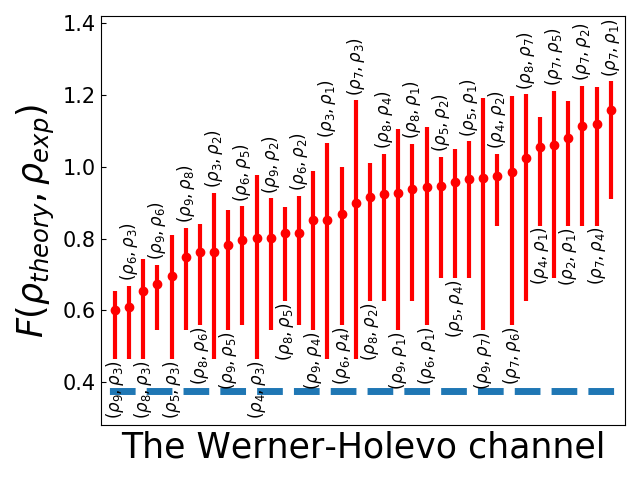}
    \end{minipage}
\caption{
    Fidelity by pairs of density matrices. Each bar shows maximum, minimum, and average (point).
    The dashed line represents the fidelity between theoretical and experimental Choi matrices $\Omega$.
}
\label{fig:more_direct_fidelity}
\end{center}
\end{figure}

In contrast to this more involved procedure, we built the Choi matrix straight forwardly using Eq.~\eqref{eq:direct_choi} on a 20-qubit computer (IBM machine \textsf{tokyo}). Its connectivity map is shown in Fig.~\ref{fig:conmap_tokyo}, where we used the highlighted 6 qubits, which are sufficient for the required placement of CNOT gates.
According to this expression, we need to prepare the state $\ket{\psi_+} = \frac{1}{\sqrt{3}} \sum_{i=1}^3 \ket{i} \ket{i}$, for which we first
create the state $\ket{\psi_{+s}} = \frac{1}{\sqrt{3}} \sum_{i=1}^3 \ket{i}$  on our system qubits using the scheme of Fig.~\ref{fig:superposition}.
Then, we place CNOT gates between system qubits and auxiliary qubits such that the control qubit is a system qubit and the target qubit is auxiliary.
After these steps, we get the state $\ket{\psi_{+}} = \frac{1}{\sqrt{3}} \sum_{i=1}^3 \ket{i} \ket{i}$ and apply the Landau-Streater (Werner-Holevo) channel transformations on this state and perform  the tomography.
The results of these experiments are shown in Fig.~\ref{fig:more_direct_fidelity}.

\section{Conclusion}
We developed and implemented algorithms for the Landau-Streater and Werner-Holevo channels.
Experiments were carried out on the 5-qubit and 20-qubit quantum computers.
The large errors encountered in the calculations can be mostly attributed to the CNOT gate errors, which have the largest error rate and are extensively used in the algorithm.
In the future we expect to reduce the number of used CNOT gates and thus to increase the overall fidelity of the experiments.

\begin{acknowledgments}
The research was supported by the Government of the Russian
Federation (Agreement 05.Y09.21.0018), Russian Foundation for
Basic Research grants Nos. 17-02-00396A, 18-02-00642A and
18-37-20073, Foundation for the Advancement of Theoretical Physics
and Mathematics ``BASIS'', the Ministry of Science and Higher
Education of the Russian Federation (16.7162.2017/8.9 and Program
No. 0066-2019-0005 for Valiev Institute of Physics and Technology
of RAS). The work at Argonne (A.G. and V.M.V) was supported by the
U.S. Department of Energy, Office of Science, Basic Energy
Sciences, Materials Sciences and Engineering Division. The
experiments on the 20-qubit machine were performed on IBM's
quantum computer \textsf{tokyo} at Oak Ridge National Laboratory
and gratefully acknowledge access to this machine.
\end{acknowledgments}

\end{document}